\documentclass[review,3p]{elsarticle}

\usepackage{amsmath,amssymb}
\usepackage{bm}
\usepackage{graphicx} 
\usepackage{wrapfig}

\def\NAME#1#2{\caption{#2} \label{#1}}

\def\SEC#1#2{\section{#1} \label{#2}}
\def\SUB#1#2{\subsection{#1} \label{#2}}
\def\SUBSUB#1#2{\subsubsection{#1} \label{#2}}

\def\EPS08#1#2#3#4{\includegraphics[#3= #4 cm,clip]{../eps08/#1.eps} \NAME{#1}{#2}}

\newfont{\bg}{cmr10 scaled\magstep5}
\newfont{\bbg}{cmsy10 scaled\magstep5}

\newcommand{\bigzerou}{\smash{\lower2.7ex\hbox{\bg 0}}}

\usepackage{lineno}
\usepackage{amsthm}
\usepackage{subfigure}
\usepackage{epsf}
\usepackage{ascmac}
\newcommand{\argmax}{\mathop{\rm arg~max}\limits} 
\usepackage{ulem} 

\modulolinenumbers[5]

\journal{Physica A}

\begin{document}

\begin{frontmatter}

\title{
Decision-making with reference information
}


\author{Riho Kawaguchi$^*$}
\address{Department of Aeronautics and Astronautics, School of Engineering, 

The University of Tokyo, 7-3-1, Hongo, Bunkyo-ku, Tokyo, 113-8656, Japan}
\cortext[mycorrespondingauthor]{Corresponding author}
\ead{riho-k@g.ecc.u-tokyo.ac.jp}

\author{Daichi Yanagisawa}

\address{Research Center for Advanced Science and Technology, 
\\The University of Tokyo, 4-6-1, Komaba, Meguro-ku, Tokyo 153-8904, Japan}
\ead{tDaichi@mail.ecc.u-tokyo.ac.jp}

\author{Katsuhiro Nishinari}

\address{Research Center for Advanced Science and Technology, 
\\The University of Tokyo, 4-6-1, Komaba, Meguro-ku, Tokyo 153-8904, Japan}
\ead{tknishi@mail.ecc.u-tokyo.ac.jp}

\begin{abstract}
We often try to predict others' actions by obtaining supporting information that shows a preference index of surrounding people. In order to reproduce these situations, we propose a game named ``One-sided Preference Game with Reference Information (OSPG-R).'' 
We conducted experiments in which players who have similar preferences compete for objects in OSPG-R. In the experiment, we used three different types of objects: boxes, faces, and cars. Our results show that the most frequently selected object was not the most popular one.
In order to gain deeper insights into the experiment's results,
we constructed a decision-making model based on two assumptions: (1) players are rational and (2) are convinced that the other players' preference orders are
equivalent to the preference index for the group. Compared to the choice behavior of the model, the experiment's results show that there was a tendency to take risks when the objects were faces, or the priority of that particular player was low. 
\end{abstract}

\begin{keyword}
Decision-making ; Multiplayer game ; Laboratory experiment ; Game Theory ; Rational choice ; Incomplete information
\end{keyword}

\end{frontmatter}

\SEC{Introduction}{SEC:INTRO}
We face numerous decision-making situations in everyday life. It is estimated that we make 226.7 decisions each day for just food alone \cite{food}. Whether decisions are made unconsciously \cite{unconscious} or consciously, we make countless decisions from choosing what to eat to making important choices such as choosing schools, partners, and jobs. Decision-making is widely studied in Psychology \cite{psychological}, Economics \cite{economics}, Mathematics \cite{mathematics}, and Political science \cite{political}. Game Theory provides useful and powerful decision-making models and  has a developed mathematical framework in many fields. It aims to understand and explain situations in which decision makers interact with each other. It provides two types of games: two-person games \cite{twoperson} and n-person ones \cite{gametheory}. Compared to two-person games, n-person ones are much more complex, but can model more realistic situations. Since every living person is confronted with decision-making situations, it is plausible to assert that we are living in a multiplayer game world.
Game Theoretical analysis for multiplayer games \cite{multi} has developed since Neumann and Morgenstern introduced them in 1944 \cite{gametheory}. Among these n-person games, there are many decision-making problems which involve the assignment of a set of objects with finite capacities to certain agents \cite{conjecture}. Those problems are called ``Matching'' problems and model markets with two kinds of heterogeneity, such as the matching of residents to hospitals \cite{Matching_medical_students_to_pairs_of_hospitals_a_new_variation_on_a_well-known_theme}, wives to husbands \cite{twosided}, students to schools \cite{twosided}, and workers to firms \cite{The_lattice_structure_of_the_set_of_stable_matchings_with_multiple_partners}. The literature on two-sided matching was introduced by Gale and Shapley in 1962 \cite{twosided} and  two-sided games have been proposed since then \cite{shapley, crawford, demange, roth, roth_sotomayor}. 

However much of the analyses in literature, including the above-mentioned papers, assume complete information; most studies assume that preferences of all agents are common knowledge. In fact, this description is not common in most real-world matching markets. Therefore, it is important to consider games where  agents on one or both sides of the market retain private information. There are papers on matching with incomplete information investigating stable matchings, i.e., no agent on either side has an incentive to change his/her partner  \cite{Two-Sided_Matching_Decision-Making_with_Uncertain_Information_Under_Multiple_States, Stable_matching_with_incomplete_information, Two-sided_matching_with_incomplete_information_about_others'_preferences, Two-sided_matching_with_interdependent_values}. Ostrovsky introduces a model of matching where, for one side, values are identical and are common knowledge, with all uncertainty concentrated on the other side in \cite{Two-Sided_Matching_with_Common_Values}. 
Bikhchandani proposes a matching model with one-sided incomplete information in which workers and firms are matched  \cite{Stability_with_one-sided_incomplete_information}. In the model, although workers have private information about their types, firms' types are common knowledge \cite{Stability_with_one-sided_incomplete_information}. Also, there have been experimental studies done on matching mechanisms with incomplete information \cite{School_choice_and_information_An_experimental_study_on_matching_mechanisms, College_admissions_and_the_role_of_information_An_experimental_study}.

A game-theoretical approach of matching problems mainly focuses on investigating stable matching and matching mechanisms, as shown above. However, when individuals make decisions freely, it is implausible that stable matching will be achieved. Thus, in this paper, we analyze the actual decision-making process and investigate how choice behaviors interact with each other.

In this paper, we propose a new game called ``One-Sided Preference Game with Reference Information (OSPG-R).'' It is a multiplayer game wherein players with similar preferences choose one object. Players' preferences are private information, but they receive public information that shows a preference index for the group. They can draw an inference on other players' preferences from the public information. They are given a rank that shows his/her ability to get the object he/she selected, which is also a private information. By conducting an experiment, we aim to analyze the actual choice behavior of players given the ambiguous available information that shows the preferences of their competitors and how they act with different positions; actions may differ between advantageous and disadvantageous positions. 

Barbey said \cite{Decision-making_is_shaped_by_individual_differences_in_the_functional_brain_connectome} that people  might apply different strategies, consider different elements of the problem or assign value to the options differently.
Moreover, he and his coauthors' study suggests that neurobiological differences appear to be important when accounting for one's susceptibility to biases in judgment and for understanding their competence in decision-making \cite{Individual_differences_in_decision_making_competence_revealed_by_multivariate_fMRI}. In our experiment, in order to investigate how chosen behaviors vary with objects, we used three different types of objects: boxes, faces, and cars. 

Furthermore, we constructed a decision-making model in order to compare choice behaviors of the experimental results with choice behavior based on a rational decision-making process. We assumed a rational player who blindly accepts reference information and analyzed the choice behavior. 

  In Section \ref{SEC:MODEL}, OSPG-R is explained in detail. Section \ref{SEC:EXPERIMENT} describes the experimental design, and Section \ref{SEC:UTILITYMAX} presents choice behavior based on rational decision-making. Section \ref{SEC:EXPRESULT} compares experiment results with theoretical results of Section \ref{SEC:UTILITYMAX}. Finally, Section \ref{SEC:CON&DIS} provides our conclusions.

\SEC{Model}{SEC:MODEL}
\SUB{Introduction of OSPG-R}{SUB:INTROofOSPGr}

In this study, we introduce a new game named OSPG-R. In this game, $n$ players make decisions and select one object among $n$ objects by taking ``preference,'' ``popularity,'' and ``priority'' into consideration. 

Each player has a ``preference'' over objects. 
The information of ``preference'' is a private information and no player besides him/her can discover this information. In this game, every player can see the information called ``popularity.'' Popularity of  objects shows a rough indication of preference orders for all players as a group. We will explain ``popularity'' in detail in  Sec. \ref{SUBSUB:POPULARITY}. Also, another information called ``priority'' will be given to each player. Priority is another private information. The number of ``priority'' tells the player where he/she ranks in achieving the object that he/she selected, among all the players. We will explain ``priority'' in detail in Sec. \ref{SUBSUB:PRIORITY}. 

\SUB{Example $(n=3)$}{SUB:EXAMPLE3}
Before illustrating $n$ - players OSPG-R , we will give an introductory explanation of a three-player OSPG-R. Consider a situation in which Player 1, Player 2, and Player 3 are to choose one object among three different objects (Object A, Object B, and Object C) . Priority is given arbitrarily for each player as shown in Tab. \ref{Tab:PRIORITY3}. The players have strict preferences over objects as shown in Tab. \ref{Tab:PREFERENCE3}. Popularity, which is calculated by a calculation method presented in Sec. \ref{SUBSUB:POPULARITY} is shown in Tab. \ref{Tab:POPULARITY3}. Only the popularity ranking (Tab. \ref{Tab:POPULARITY3}) is common knowledge among the players.
\begin{table}[htbp]
  \centering
  \caption{Priority of each player.}

\begin{tabular}{|c||c|} \hline
 Priority&Player\\ \hline\hline
 First&Player 1\\ \hline
 Second&Player 3\\ \hline
 Third&Player 2\\ \hline
    \end{tabular}%
  \label{Tab:PRIORITY3}%
\end{table}%

\begin{table}[htbp]
  \centering
  \caption{Player's preferences over objects.}

\begin{tabular}{|c||c|c|c|} \hline
 Preference&Player 1&Player 2& Player 3\\ \hline\hline
 First&Object A&Object A&Object B\\ \hline
 Second&Object B&Object B&Object A\\ \hline
 Third&Object C&Object C&Object C\\ \hline
    \end{tabular}%
  \label{Tab:PREFERENCE3}%
\end{table}%

\begin{table}[htbp]
  \centering
  \caption{Popularity of the objects.}

\begin{tabular}{|c||c|} \hline
 Popularity&Object\\ \hline\hline
 First&Object A\\ \hline
 Second&Object B\\ \hline
 Third&Object C\\ \hline
    \end{tabular}%
  \label{Tab:POPULARITY3}%
\end{table}%

\clearpage
Here, one may imagine what their actions may be. 
We will explain two cases of choice behaviors by the players. Note that these are only examples and that they are to illustrate how OSPG-R is played.
\begin{description}
 \item[Case 1:]Player 1 may choose Object A because his/her priority is first and he/she may thus obtain any object he/she desires. Player 2 may assume that other players' preferences would be same as the popularity ranking and they would choose Object A or Object B, so his/her chance of obtaining them would be small. Nevertheless he/she may insist on choosing Object B. Player 3 may guess that there may be a high possibility of obtaining Object B taking his/her priority, which is the second among all the players and popularity, implying that Object B is second most popular into account. 
Based on the situation above, choice behaviors of the players would be as shown in Tab. \ref{Tab:CHOICE31}.
Then, the results of the game would be as shown in Tab. \ref{Tab:RESULT31}. Because Player 1 has the highest priority, he/she gets any object he/she wants: Player 1 gets Object A. Player 3 has the second highest priority and his/her choice Object B differs from Player 1's, so his/her choice goes through. On the other hand, Player 2 gets nothing because he/she has the lowest priority and his/her choice of Object B is same as Player 3's choice. 
\begin{table}[htbp]
  \centering
  \caption{Choice behavior of each player (Case 1).}

\begin{tabular}{|c||c|c|c|} \hline
Player&Player 1&Player 2&Player 3\\ \hline\hline
Choice&Object A&Object B&Object B\\ \hline
    \end{tabular}%
  \label{Tab:CHOICE31}%
\end{table}%

\begin{table}[htbp]
  \centering
  \caption{Results of the game (Case 1).}

\begin{tabular}{|c||c|c|c|} \hline
Player&Player 1&Player 2&Player 3\\ \hline\hline
Results&Object A&Nothing&Object B\\ \hline
    \end{tabular}%
  \label{Tab:RESULT31}%
\end{table}%
\clearpage
 \item[Case 2:]Player 1 may choose Object A because his/her priority is first and he/she may claim any object he/she desires. Player 2 may assume that other players' preferences would be same as the popularity ranking and they would choose Object A or Object B. Yet, because he/she has little opportunity to get Object A or Object B, he/she may choose Object C. Player 3 is skeptical about the popularity ranking and predicts that the highest priority player would choose Object B and he/she has no chance to get it. Therefore, he/she may choose Object A.
 Based on the type of situation above, choice behavior of the players would be as shown in Tab. \ref{Tab:CHOICE32}.
Then, the results of the game would be as shown in Tab. \ref{Tab:RESULT32}. Because Player 1 has the highest priority, he/she gets any object he/she wants: Player 1 gets Object A. Player 3 gets nothing because his/her choice is same as Player 1's. Player 2 gets Object C.
\end{description}

\begin{table}[htbp]
  \centering
  \caption{Choice behavior of each player (Case 2).}

\begin{tabular}{|c||c|c|c|} \hline
Player&Player 1&Player 2&Player 3\\ \hline\hline
Choice&Object A&Object C&Object A\\ \hline
    \end{tabular}%
  \label{Tab:CHOICE32}%
\end{table}%

\begin{table}[htbp]
  \centering
  \caption{Results of the game (Case 2).}

\begin{tabular}{|c||c|c|c|} \hline
Player&Player 1&Player 2&Player 3\\ \hline\hline
Results&Object A&Object C&Nothing\\ \hline
    \end{tabular}%
  \label{Tab:RESULT32}%
\end{table}%

\SUB{OSPG-R}{SUB:OSPGr}

Let us consider decision-making in a  simultaneous game of $n\ (\geq2) $ players choosing between $n$ objects.
 Each player $ \tilde{i}$ has his/her preference $\tilde{p}_{ \tilde{i}\tilde{j}}$ over object $ \tilde{j}$ and it is represented by ranking without ties. Let $N_{\rm p} =\{1, 2, \ldots,\tilde{i}, \ldots, n \}$ denote a set of players and $N_{\rm o} =\{1, 2, \ldots, \tilde{j}, \ldots,n \}$ denote a set of objects. Players receive information on the objects' popularity $\tilde{q}_{\tilde{j}}$ (see Sec. \ref{SUBSUB:POPULARITY}). Each player is given a priority number $\tilde{r}_{ \tilde{i}}$ (see Sec. \ref{SUBSUB:PRIORITY}). Note that $\tilde{q}_{\tilde{j}}$ is the only public information available to all players. 
 
 Each player chooses one object by taking up his/her preference, his/her priority, and the popularity of objects into account. The game result, whether a player attains the object he/she chose, is decided by his/her priority $\tilde{r}_{ \tilde{i}}$ (see Sec. \ref{SUBSUB:PRIORITY}). The player with the highest priority obtains the object he/she desires and the others get nothing. If no player except player $\tilde{i}$ chooses the same object, he/she gets the object he/she chooses. On the other hand, if player $\tilde{i}$ and player $-\tilde{i}\in N-\{\tilde{i}\}$ choose the same object, player $\tilde{i}$ gets the object when $\tilde{r}_{ \tilde{i}}<\tilde{r}_{ -\tilde{i}}$.

\SUBSUB{Definition of Popularity}
{SUBSUB:POPULARITY}

For each object $\tilde{j}$, we have $\tilde{p}_{ \tilde{i}\tilde{j}}$ denote the preference rank order placed by player $\tilde{i}$.
 Score $s_{\tilde{j}}$ for object $\tilde{j}$ is calculated by Borda score $s_{\tilde{j}}=\sum ^{n}_{i=1} (n-(\tilde{p}_{ \tilde{i}\tilde{j}}-1) )$ \cite{borda}. We define $\tilde{q}_{ \tilde{j}} \in \{1, 2, \ldots, n\},\ \tilde{q}_{ \tilde{j}}\neq \tilde{q}_{- \tilde{j}}$,  $-\tilde{j} \in N_{\rm o}-\{\tilde{j}\}$  as popularity ranking of object $\tilde{j}$. $\tilde{q}_{ \tilde{j}}=k$ holds when  $s_{\tilde{j}}$ is the $k$ th largest score among $S=\{s_1, \ldots, s_{\tilde{j}}, \ldots, n\}$. The larger the score $s_{\tilde{j}}$ for the object is, the higher the popularity is.

\SUBSUB{Definition of Priority}{SUBSUB:PRIORITY}

A priority indicates at what number a player can attain an object that he/she chooses, among all the players. A priority number $\tilde{r}_{ \tilde{i}} \in \{1, 2, \ldots, n\},\ \tilde{r}_{ \tilde{i}}\neq \tilde{r}_{ -\tilde{i}}$, $-\tilde{i} \in N-\{\tilde{i}\}$ is an ordering of  player $\tilde{i}$ in $N_{\rm p}$. For example, a player given $\tilde{r}_{\tilde{i}}=1$ has the first priority to attain an object, $\tilde{r}_{ \tilde{i}}=2$ has the second priority to attain it, and player given $\tilde{r}_{ \tilde{i}}=n$ has the lowest priority. In other words, if plural players chose the same object, the player with the highest priority among those players obtains it. \\

For the following discussions, $i=1, 2, \ldots,n$ will be cited for player $\tilde{i}$ whose priority number satisfies $\tilde{r}_{ \tilde{i}}=i$ and $j=1, 2, \ldots,n$ will be cited for object $\tilde{j}$ whose popularity ranking satisfies $\tilde{q}_{ \tilde{j}}=j$ without loss of generality. Also, preference $\tilde{p}_{ \tilde{i}\tilde{j}}$ of player $i$ over object $j$ will be cited as $p_{ {i}{j}}$ without loss of generality.

\SEC{Experiment}{SEC:EXPERIMENT}

\SUB{Experimental design}{SUB:DESIGN}
We conducted experiments in which groups of five players played in OSPG-R with three different objects: pictures of boxes without characters, pictures of people's faces of a different gender, and pictures of cars. There are five objects: A, B, C, D, and E for each type of object. We chose boxes without characters other than numbers from 1 to 5, written on them in order to unify the players' utility for obtaining a particular box.
However, we selected faces as objects, because they contrast with boxes in the sense of utility. We assumed that utility for faces would differ largely among players and it would affect the players' choice behaviors. We expected cars to be the intermediate object between boxes and faces. 

Thirty students from The University of Tokyo were recruited to participate in the OSPG-R experiment. All students were paid a flat fee for participation in this experiment. First, players submitted preference orders over faces and cars. Since boxes are without any character, we had the players to write the same preference order over box A to E, as face A to E. 

We divided thirty players into six groups of five players, so as Kendall tau distance \cite{kendall} $\tau_i$ (see Appendix \ref{SEC:tau}) between preference ranking of every player $i$ and popularity raking of a group is to be smaller than three. The smaller the $\tau_i$ is, the more the preference of a player is similar to popularity. In fact, one group when choosing over boxes and faces, and two groups when choosing over cars, did not satisfy this condition. 
Thus, the result of the other five and four groups were noteworthy in the case of boxes and faces and in the case of cars, respectively.

After being divided into groups, players received information on popularity ranking of the objects of their group. Players independently chose one object among five in each round in which players received a priority number. There were five rounds in each group and the priority of players changed after every round. Thus, the players experienced every priority, 1 to 5, through the five rounds. A player was told to choose the object based on his/her preference that he/she submitted in advance. In order to reduce the learning effect to a minimum, players did not receive any feedback about previous choices or outcomes. 

OSPG-R with object boxes, faces, and cars was played in that order. In game with boxes, players were told to assume there were prizes in the box corresponding to the preference order which is equivalent to that of the faces case. For example, if a player's preference order for Box C was 2, second prize was in Box C. 

Although conditions for rewards were identified in induced value theory \cite{inducedvaluetheory} from Experimental Economics, we did not give rewards according to the player's result from the game in order to prevent players from acting by some behavioral principle of pecuniary motive, because we intended to investigate the difference between types of objects. 



\SEC{Rational decision-making with reference information (RDM-R)}{SEC:UTILITYMAX}

The decision-making process is widely studied in Game Theory \cite{gametheory}. Lucas says that Game theory is  ``a collection of mathematical models formulated to study decision-making in situations involving conflict and cooperation $\ldots$ It is concerned with finding optimal solutions or stable outcomes when various decision makers have conflicting objectives in mind.'' \cite{lucas} 
In game theory, rational self-interested players maximize their outcomes. Although there are criticisms that those characteristics deviate from decision-making in reality,  Game Theory is a useful means to model interaction between decision makers. 

In this section, we explain the choice behavior for a player with rational decision-making, yet who blindly accepts the reference information popularity. We will call the aforementioned decision-making process ``RDM-R,'' defined as follows:
\begin{description}
\item[Assumption 1:] Players are rational agents \cite{rational}.
\item[Assumption 2:] A player is convinced that the preferences of others and popularity ranking are the same.
\end{description}
 Such players assume that other players have preferences that are the same as popularity, because it is difficult for players to predict each other's preferences. 

We explain the choice behavior based on RDM-R.
In this game, players rank their preferences in a constant way as mentioned in Sec. \ref{SUB:INTROofOSPGr}. Preferences satisfy the axioms of transitivity, completeness, and continuity. Therefore, by the theorems of Debreu \cite{debreu}, an ordinal utility function exists.
We define a utility function for each object as $u_{ij}=n+1-p_{ij}$.

Player $i$ who has a decision-making process based on the assumptions above assumes that player $-i (< i)$ who has higher priority chooses object $j=i$ (see Appendix \ref{SEC:proof1}). Therefore, payoff for each player $i=1, 2, \ldots, n$ to choose object $j=1, 2, \ldots, n$ can be written as shown in Tab. \ref{Tab:PAYOFFplayer-object}.

\begin{table}[htbp]
  \centering
  \caption{Payoff for each player $i=1, 2, \ldots, n$ to choose object $j=1, 2, \ldots, n$.}

\begin{tabular}{|c|c|c|c|c|c|c|c|} \hline
\multicolumn{2}{|c|}{} & \multicolumn{6}{c|}{object }\\ \cline{3-7}
\multicolumn{2}{|c|}{} &$1$ & $2$ &$3$& $\cdots$ & $n$ \\ \hline
&$1$& $u_{11}$ & $u_{12}$ &$u_{13}$& $\cdots$ & $u_{1n}$ \\ \cline{2-7}
&$2$& $0$ & $u_{22}$ &$u_{23}$& $\cdots$  & $u_{2n}$    \\ \cline{2-7}
player&$3$& $0$ & $0$ &$u_{33}$& $\cdots$  & $u_{3n}$    \\ \cline{2-7}
&$\vdots$& $\cdots$& $\cdots$ &$\cdots$& $\cdots$ & $\cdots$    \\ \cline{2-7}
&$n$&$0$ & $0$ &$0$& $\cdots$ &  $u_{nn}$ \\ \hline
    \end{tabular}%
  \label{Tab:PAYOFFplayer-object}%
\end{table}%

Tab. \ref{Tab:PAYOFFplayer-object} shows that it is rational for player $i$ to choose object $\argmax_{j} \ u_{ij} , j \in N_{\rm o} -\{ j=1, 2,  \ldots, i-1\}$ (see Appendix \ref{SEC:proof2}).

\SEC{Experimental results and theoretical results}{SEC:EXPRESULT}

\SUB{Comparison of choice behavior between experimental result and RDM-R}{SUB:COMPARISON}
 Since each player chose one object given different priority at every round and decision-making was independent during each round, we were able to reform $5!=120$ new groups from one group by extracting five players with five different priorities from a group formed in Sec. \ref {SUB:DESIGN}. 
 
 Therefore, we obtained 120 (priority patterns) $\times$ 5 (popularity patterns) data for the faces and boxes cases, and 120 (priority patterns) $\times$ 4 (popularity patterns) data for the cars case.
 
 Figure \ref{Fig:modboxwomcar} shows the average rate of chosen times. Fig. 1A shows the average result of five groups with different popularity in boxes and faces case (average $\tau_i=0.96$). Fig. 1B shows average result of four groups with different popularity patterns in cars case (average $\tau_i=1.05$). RDM-R is the theoretical result, and a box, woman, or car are experimental results.
 
 As we can see in both figures, the most popular object, Object $1$, was not the object which was most frequently selected. What is of interest is that although Object 1 was the most popular object, when choice behavior was observed macroscopically, it was not chosen by the majority. 
 
  Moreover, as long as the $\tau_i$ of all players is less than 3, the same is true for average choice behavior of  every group with different popularity pattern based on RDM-R. Fig. \ref{Fig:mod} is choice behavior of every possible combination for every preference, which satisfies (A) $\tau_i=0$, (B) $\tau_i \le 1$, and (C) $\tau_i \le 2$. As the range of $\tau_i$ increases, the rate of times that Object $1$ is chosen decreases and the rate of times that Object $5$ is chosen increases. Interestingly, the object which has the lowest priority is chosen the most when choice behavior is based on RDM-R.
 \begin{figure}[htbp]
  \begin{center}
\begin{minipage}{.45\linewidth}  
 \includegraphics[width=1.1 \linewidth]{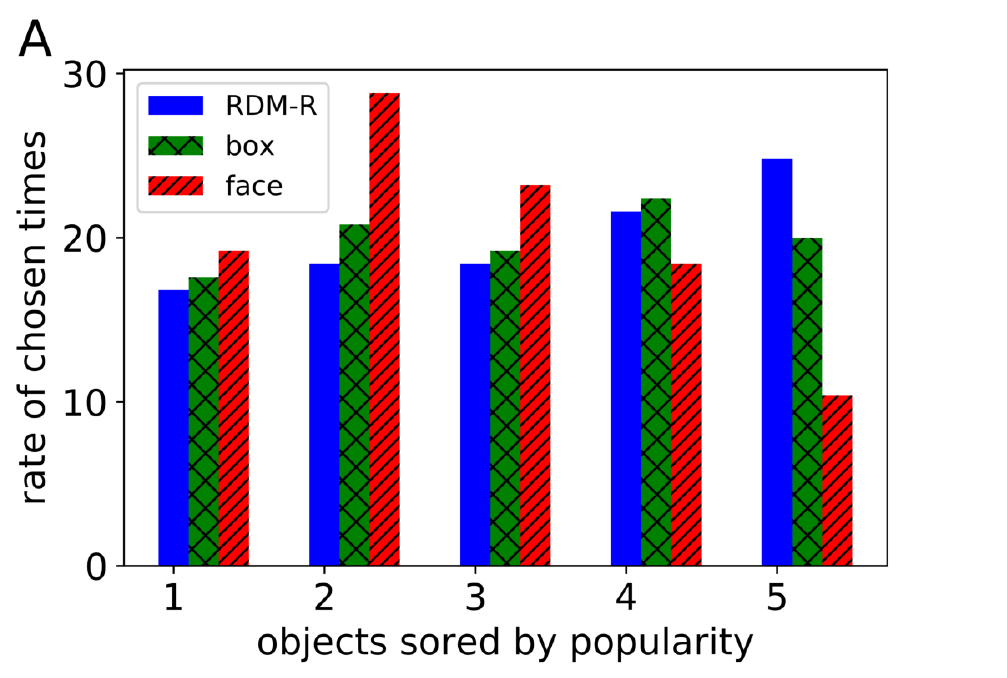}
   \end{minipage}
  \hspace{.5pc}
  \begin{minipage}{.45\linewidth} 
  \includegraphics[width=1.1 \linewidth]{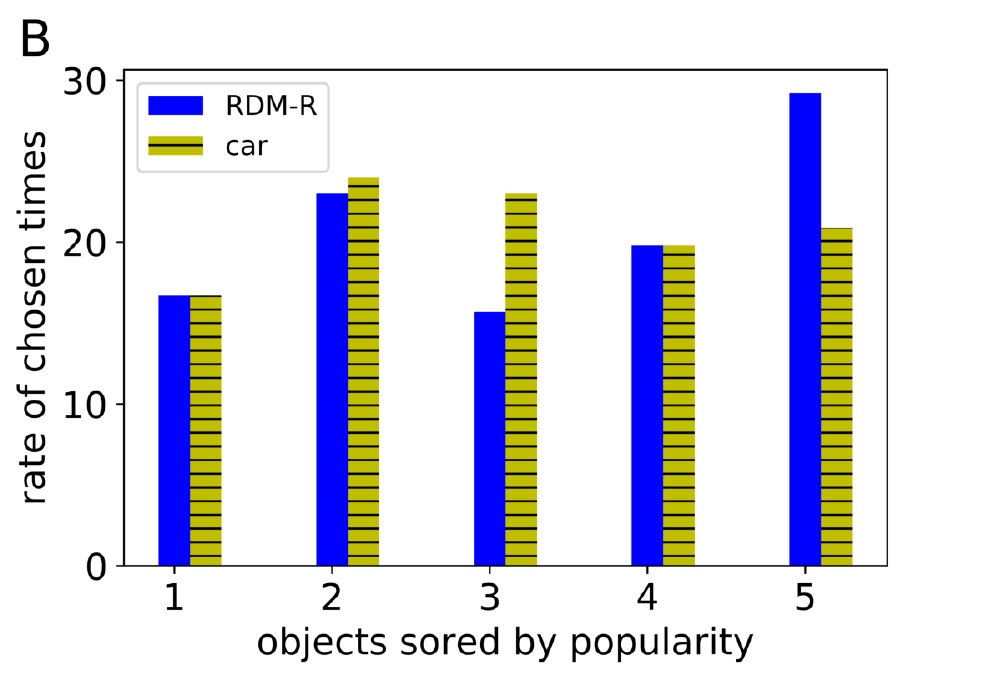}\label{Fig:modboxwomcar}
   \end{minipage}
  \end{center}
\caption{(A) The average rate of chosen times of each object  when the experimental preferences for boxes and faces was used. Blue, green, and red bars represent the results of RDM-R, boxes, and faces, respectively. (B) The average rate of chosen times of each object  when the experimental preferences for cars were used. Blue and yellow bars represent the results of RDM-R and cars, respectively.}
\label{Fig:modboxwomcar}
\end{figure}

 \begin{figure}[htbp]
  \begin{center}
\begin{minipage}{.3\linewidth}  
 \includegraphics[width=1.2 \linewidth]{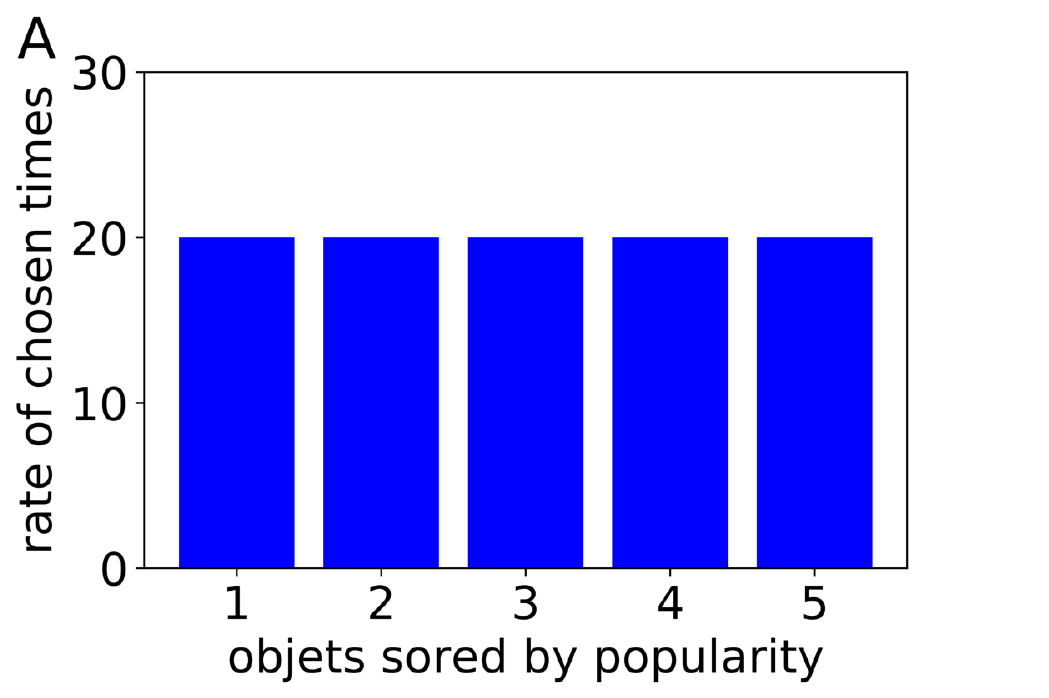}
   \end{minipage}
  \hspace{.5pc}
  \begin{minipage}{.3\linewidth} 
  \includegraphics[width=1.2 \linewidth]{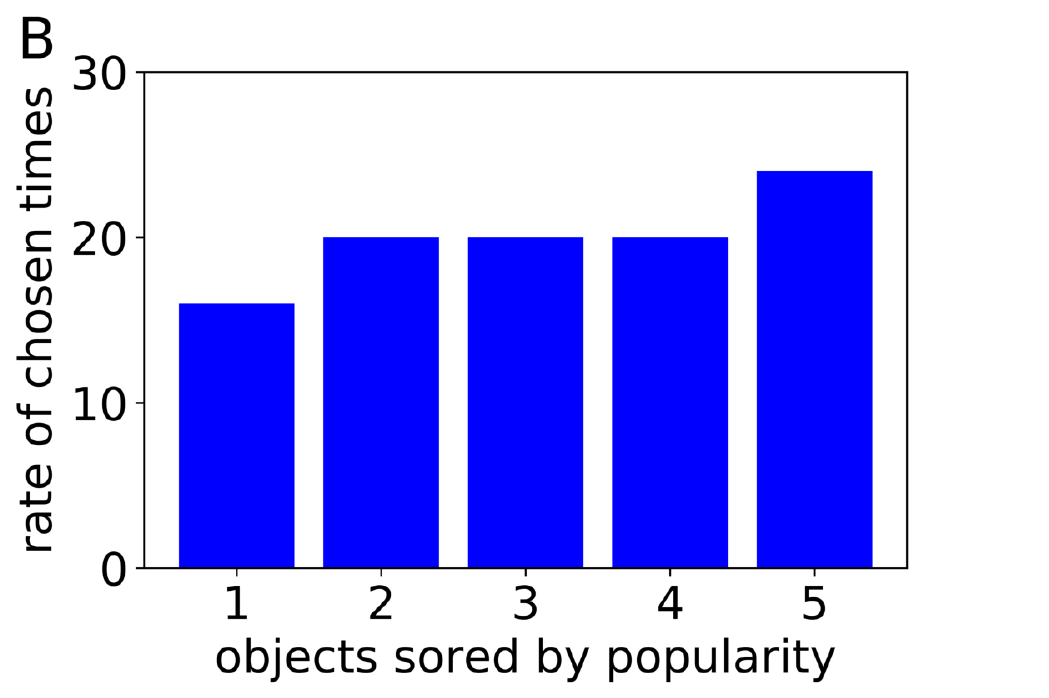}
   \end{minipage}
    \hspace{.5pc}
  \begin{minipage}{.3\linewidth}  
  \includegraphics[width=1.2 \linewidth]{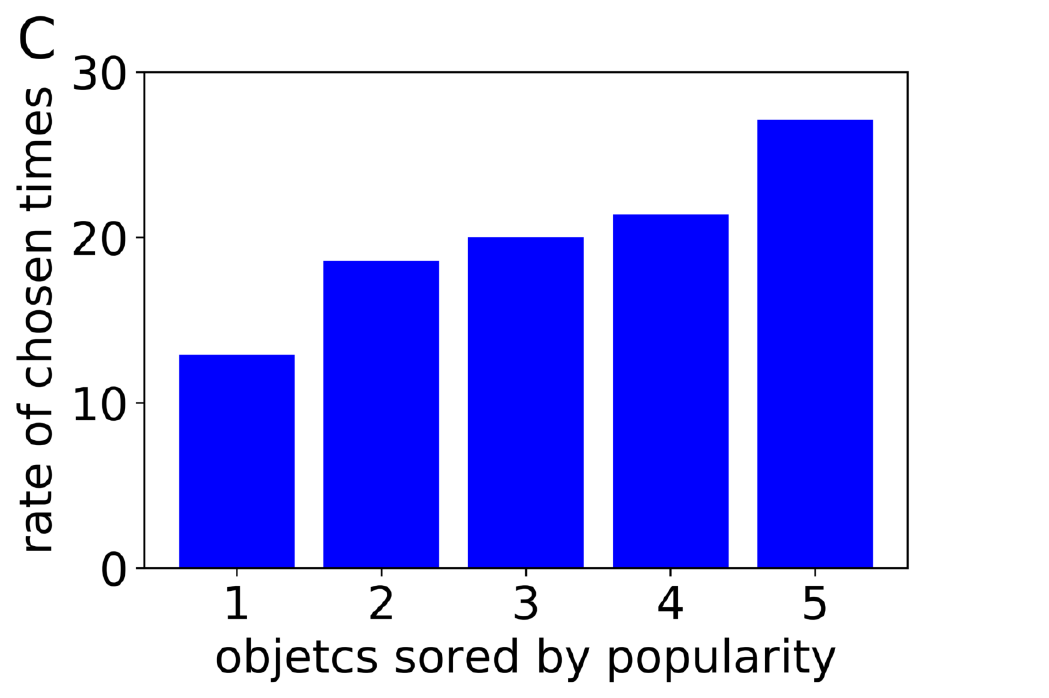}
   \end{minipage}
  \end{center}
  \caption{The rate of chosen times by RDM-R for each object of every combination of each preference. (A) $\tau_i=0$, (B) $\tau_i \le 1$, and (C) $\tau_i \le 2$.}
\label{Fig:mod}
\end{figure}
\SUB{Deviation of choice behavior between experimental result and RDM-R}{SUB:DEVIATION}

As shown in Fig. \ref{Fig:modboxwomcar}, there was deviation between the choice behaviors of players and the decision-making process based on RDM-R.
In this subsection, we introduce three terms in order to define deviation as follows.

\begin{enumerate}
\item RDM-R
\item Risk
\item Safe
\end{enumerate}
If a player chose an object following RDM-R, we refer to the choice as ``RDM-R.'' If a player chose an object which was  preferred to the object which was chosen by RDM-R, we refer to the choice as ``Risk.'' If a player chose an object which was not preferred to the object which was chosen by RDM-R, we refer to the choice as ``Safe.'' 
In Sec. \ref {SUBSUB:COMfocusOBJ}, we focus on the types of the objects. Then, in Sec. \ref {SUBSUB:COMfocusPRI}, we analyze the results more in detail by considering the priority.

\begin{figure}[htbp]
\centering
\includegraphics[width=8cm]{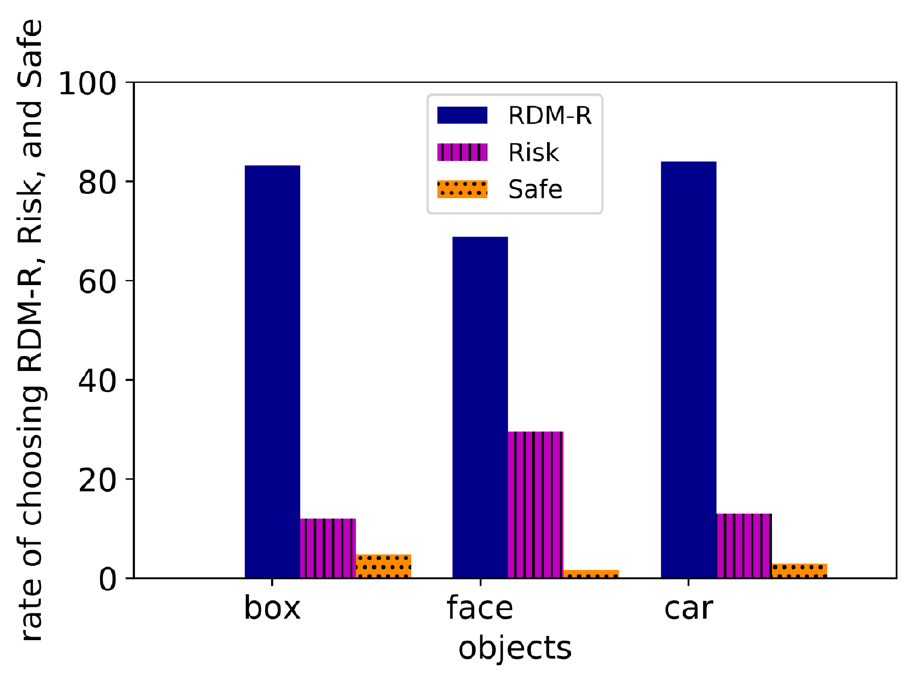}
\caption{Rate of  choice of RDM-R, Risk, and Safe by the players.}
\label{Fig:object}
\end{figure}

\SUBSUB{Comparison of experimental results focusing on types of objects}{SUBSUB:COMfocusOBJ}
Figure \ref{Fig:object} shows rates of choices of RDM-R, Risk, and Safe by the players. As Fig. \ref{Fig:object} shows, choices by players being the same as choice of RDM-R decreased greatly when the objects were faces. On the one hand, there was a tendency for the players to choose Risk when the objects were faces rather than when the objects were boxes and cars. On the other hand, choices of Safe did not increase when the objects were faces. We assume that the reason why many players chose Risk when the objects were faces is that utility differed largely with players and that the difference between utility for preferable objects and utility for unpreferable ones was large. Players may have thought that it was better to take risks by choosing more preferable objects rather than choosing those other ones. However, when the objects were boxes and cars, utility did not differ largely with players and that the difference between utility for preferable objects and utility for unpreferable ones was not large. Therefore, players may have thought that it was better than getting nothing and thus are less likely to take risks. The rate of choices of Risk when objects were cars did not differ greatly from those of boxes, contrary to our expectations that the choice behavior for cars become intermediate between those of boxes and faces. 
From this result, we considered that small difference of utilities among objects did not affect the choice behaviors such as boxes and cars. On the contrary, choice behaviors for faces greatly differed from those for the other two objects because of the large difference of utility.

\begin{figure}[htbp]
\centering
\includegraphics[width=8cm]{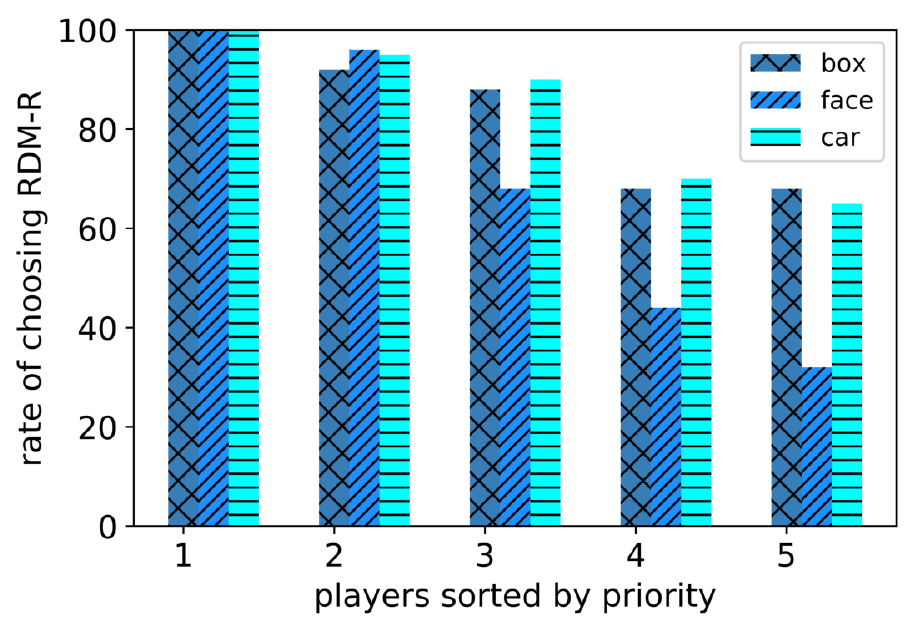}
\caption{Rates of choices of RDM-R.}
\label{Fig:prioritymod}
\end{figure}

\begin{figure}[htbp]
\centering
\includegraphics[width=8cm]{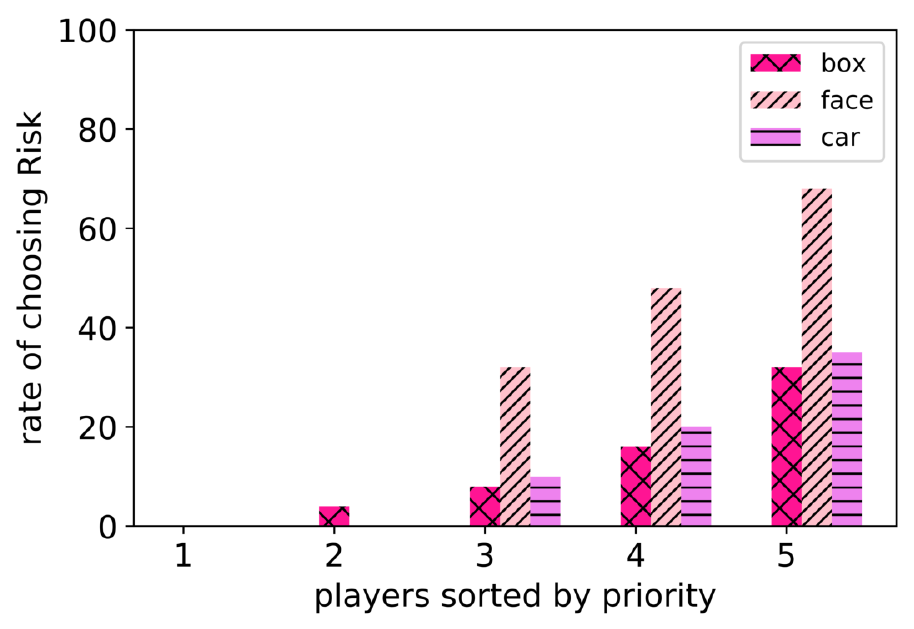}
\caption{Rates of choices of Risk.}
\label{Fig:priorityrisk}
\end{figure}

\begin{figure}[htbp]
\centering
\includegraphics[width=8cm]{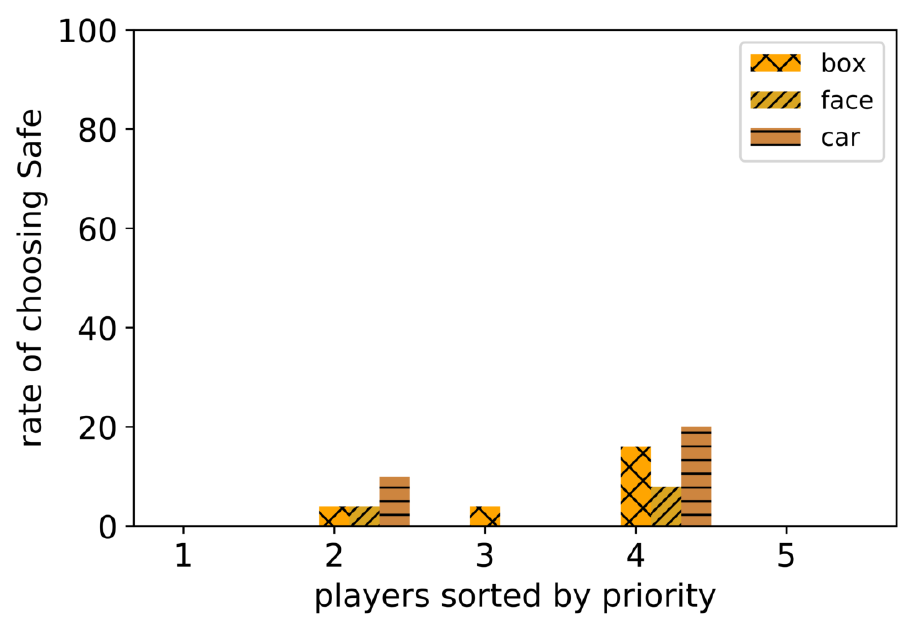}
\caption{Rates of choices of Safe.}
\label{Fig:prioritysafe}
\end{figure}

\SUBSUB{Comparison of experimental results focusing on priority of the players}{SUBSUB:COMfocusPRI} 
Figure \ref{Fig:prioritymod},  \ref{Fig:priorityrisk}, and \ref{Fig:prioritysafe} shows the rates of choices of RDM-R, Risk, and Safe for each type of object. As seen in Fig. \ref{Fig:prioritymod}, when a player had the highest priority, he/she certainly chose the same object chosen by RDM-R. The lower the priorities of the players were, the less frequent were the choices of RDM-R. Moreover, compared to the results for boxes and cars, the results for faces deviated largely as priorities become low. 

Figure \ref{Fig:priorityrisk} shows that the lower priorities of players were, the more the choices for Risk increased. The increase of taking risks was specific when the objects were faces. We considered the reason why was because players did not have many choices with high possibility of obtaining the object when their priorities were low, and may have had negative utility for some faces, and thus they took risks.

As shown in Fig. \ref{Fig:prioritysafe}, the rate of choosing Safe was zero for players who had the highest priority and the lowest priority. If choice behavior of players had deviated randomly from RDM-R, the distribution of choosing Risk and Safe would have been the same, because Risk and Safe are symmetrical. However, interestingly, players tended to choose Risk rather than Safe. Thus, our result suggests that people are likely to take risks especially when their priorities are low in decision-making situations.
Thus, the deviation of the choice behavior when objects were faces mainly resulted from the choice behaviors of players whose priorities were low.

\clearpage
\SEC{Conclusions}{SEC:CON&DIS}
In this paper, we proposed OSPG-R, a multiplayer decision-making game with reference information, and analyzed choice behaviors of experimental results and a rational decision-making process. Although many studies of matching mechanisms assume complete information, it is an implausible assumption because obtaining information of every agent in the market and dealing with all the information is unrealistic. In order to deal with this problem, we provided some reference information to the players. We intended to examine how players' choice behaviors would appear in a group from a macroscopic view point. We also analyzed the choice behaviors of players by comparing them to those of rational decision-making actors.

First, both experimental results and theoretical results show that the most popular object, i.e., the object that is considered to be the most attractive by the majority, is not selected most often in the context of decision-making by players with similar preferences. The result indicates that considering popularity and their priorities, players compromise their preferences. As a result, players tended to avoid choosing the most popular object.

Secondly, there is deviation between experimental results and theoretical results. Choice behavior differed distinctively by objects: boxes, faces, and cars. Our intention was to investigate how choice behavior differs with objects. When the objects were faces, players' choice behaviors deviated widely from rational decision-making model since they tended to take risk. We considered that this action was due to the fact that players take risks when the objects are faces, because utility is not unified and some may think that it is better to take a risk by choosing more preferable objects rather than choosing less-preferable ones. Furthermore, the lower the priorities of players were, the fewer choices of rational decision-making were available and, consequently more risk-taking choices were undertaken. We considered that the reasons for this outcome had to do with the fact that players do not have many choices with high possibility of obtaining the object when their priorities are low and they may only gain significantly low utility for some objects; Therefore, he/she undertakes risky behavior.

We believe that this research can be extended to mimic real situations within a matching market and can be used to understand how people react to available information, and therefore contribute to predict crowd mentality with controlled information in future studies.

\section*{Acknowledgment}
This work was supported by JST-Mirai Program Grant Number JPMJMI17D4,
Japan and JSPS KAKENHI Grant Number JP15K17583.

The experiments in this paper have been reviewed and approved by the Research Ethics Committee at The University of Tokyo, Japan (No. 17-158).


\appendix
\SEC{Kendall tau distance $\tau_i$}{SEC:tau}
We explain Kendall tau distance in this Appendix.

For each object $j$, player $i$ ranks $p_{ij} $. Kendall tau distance is an inversion number \cite{flip}, which is the smallest number of exchanges needed to sort the sequence of preference order $ (p_{i1} \ \ldots\ p_{ij}\ \ldots\ p_{in})$. The larger $\tau_i$ is, the more different  from popularity preference of player is. 

For example, in the case of Sec. \ref{SUB:EXAMPLE3}, Kendall tau distance $\tau_i$ of Player A, B, and C can be calculated as noted in Tab. \ref{Tab:FLIPPING3}. Since the preference sequence of players A and B is $(1 \ 2\ 3)$ and the popularity sequence is $(1 \ 2\ 3)$, $\tau_{\rm A}=\tau_{\rm B}=0$. Yet, it takes one exchange, $2$ and $1$, to sort Player C's preference sequence $(2 \ 1\ 3)$. Therefore, $\tau_{\rm C}=1$.

\begin{table}[htbp]
  \centering
  \caption{$\tau_i$ of each player's preference order over objects.}

\begin{tabular}{|c||c|c|c|} \hline
Player&Player A&Player B&Player C\\ \hline\hline
$\tau_i$&0&0&1\\ \hline
    \end{tabular}%
  \label{Tab:FLIPPING3}%
\end{table}%
\clearpage


\SEC{Proof 1}{SEC:proof1}

We prove that choosing object $j=i$ is a rationalizable strategy for player $i ( \forall i \in N-\{1\} )$ whose preference is equivalent to popularity with a decision-making process of RDM-R.
\begin{proof}
$  \ \ \ \ \ \  \ \ \ \ \ \ \ \ \ \ \ \ \ \ \ \ \ \ \ \ \  \ \ \ \ \ \ \ \ \ \ \ \ \ \ \ \ \ \ \ \ \ \ \ \ \ \ \ \  \ \ \ \  \ \ \ \ \ \  \ \ \ \ \ \ \ \ \ \ \ \ \ \ \ \ \ \ \ \ \  \ \ \ \ \ \ \ \ \ \ \ \ \ \ \ \ \ \ \ \ \ \ \ \ \ \ \ \  \ \ \ \  \ \ \ \ \ \ \ \ \ \ \ \  \ \ \ \ \ \ \ \  \ \ \ \ \ \ \ \ \ \ \ \ \ \ \ \  \ \ \ \ \ \ \ \ \ \ \ \ \ \ \ \ \ \ \ $
Initial step: fix player $i=1$.
player $1$'s dominant strategy is choosing object $1$, because he/she has the highest popularity and he/she prefers object $1$ most, which is clearly true.$  \ \ \ \ \ \  \ \ \ \ \ \ \ \ \ \ \ \ \ \ \ \ \ \ \ \ \  \ \ \ \ \ \ \ \ \ \ \ \ \ \ \ \ \ \ \ \ \ \ \ \ \ \ \ \  \ \ \ \  \ \ \ \ \ \ \ \ \ \ \ \  \ \ \ \ \ \ \ \  \ \ \ \ \ \ \ \ \ \ \ \ \ \ \ \  \ \ \ \ \ \ \ \ \ \ \ \ \ \ \ \ \ \ \ $
Inductive Step: Suppose that the rationalizable strategy for each player $i=1, 2, \ldots, k-1 $ is choosing object $j=1, 2, \ldots, k-1$. It remains to show that rationalizable strategy for player $k$ is choosing object $k$. Since player $k$ knows that rationalizable strategy for each player $i=1, 2, \ldots, k -1$ is choosing object $j=1, 2, \ldots, k-1$, every payoff for choosing object $j=1, 2, \ldots, k-1$ is $0$. The same thing can be said for player $k+1, k+2, \ldots, n$. Tab. \ref{Tab:PAYOFFk} provides payoff matrix of player $i=k$ and player $i=k'(>k)$. For example, for cell $(n+1-k, 0)$, the first number is the payoff to player $k$ when he/she chooses object $k$ and the second number is the payoff to player $k'$ when he/she chooses object $1$.  It shows that choosing object $k$ is the best response for player $k$.
Therefore, the rationalizable strategy for player $k$ is to choose object $k$.
Thus, by the principle of mathematical induction, the hypothesis is proven.
 \begin{table}[htbp]
  \centering
  \caption{Payoff matrix of player $k$ and player $k'$.}

\begin{tabular}{|c|c|c|c|c|c|c|c|c|} \hline
\multicolumn{2}{|c|}{} & \multicolumn{6}{c|}{player $k'$}\\ \cline{3-8}
\multicolumn{2}{|c|}{} &$1$ & $\cdots$ &$ k-1$ & $k$ & $\cdots$ & $n$ \\ \hline
&$1$& $(0, 0)$ & $\cdots$ & $(0,0)$ & $(0, 0)$ & $\cdots$ & $(0, 1)$ \\ \cline{2-8}
&$\vdots$& $\cdots$& $\cdots$ & $\cdots$ & $\cdots$ & $\cdots$ & $\cdots$    \\ \cline{2-8}
player $k$&$k-1$& $(0, 0)$ & $\cdots$ & $(0,0)$ & $(0, 0)$ & $\cdots$ & $(0, 1)$ \\ \cline{2-8}
&$k$&$(n+1-k, 0)$ &  $\cdots$  & $(n+1-k, 0)$ & $(n+1-k, 0)$ & $\cdots$ & $(n+1-k, 1)$    \\ \cline{2-8}
&$\vdots$&$\cdots$& $\cdots$ & $\cdots$ & $\cdots$ & $\cdots$ & $\cdots$    \\ \cline{2-8}
&$n$&$(1, 0)$ & $\cdots$ & $(1, 0)$ & $(1, n+1-k)$ & $\cdots$ & $(1, 0)$ \\ \hline
    \end{tabular}
  \label{Tab:PAYOFFk}%
\end{table}%
\end{proof}

\clearpage

\SEC{Proof 2}{SEC:proof2}
We prove that rationalizable strategy for player $i$ is to choose object $\argmax_{j} \ u_{ij}, j \in N_{\rm o} -\{ j=1, 2,  \ldots, i-1\}$.

\begin{proof}$  \ \ \ \ \ \  \ \ \ \ \ \ \ \ \ \ \ \ \ \ \ \ \ \ \ \ \  \ \ \ \ \ \ \ \ \ \ \ \ \ \ \ \ \ \ \ \ \ \ \ \ \ \ \ \  \ \ \ \  \ \ \ \ \ \  \ \ \ \ \ \ \ \ \ \ \ \ \ \ \ \ \ \ \ \ \  \ \ \ \ \ \ \ \ \ \ \ \ \ \ \ \ \ \ \ \ \ \ \ \ \ \ \ \  \ \ \ \  \ \ \ \ \ \ \ \ \ \ \ \  \ \ \ \ \ \ \ \  \ \ \ \ \ \ \ \ \ \ \ \ \ \ \ \  \ \ \ \ \ \ \ \ \ \ \ \ \ \ \ \ \ \ \ $
Based on Assumption 2 of RDM-R described in Sec. \ref{SEC:UTILITYMAX}, player $-i$'s preference order is in the sequence of $1, 2, \ldots,  j, \ldots,n $, so utility for each object is $n, n-1, \ldots, n+1-j, \ldots,1$. Based on Assumption 1, choosing object $j=-i$ is proved in Appendix \ref {SEC:proof1} to be a rationalizable strategy for player $-i$ whose preference is equivalent to popularity with a decision-making process of RDM-R. His/Her payoff for choosing object $j=1, 2, \ldots ,i-1$ is $0$. Tab. \ref{Tab:PAYOFFi} is player $i$'s payoff against player $i'(>i)$. Therefore, the rationalizable strategy for player $i$ is to choose object $\argmax_{j} \ u_{ij}, j \in N_{\rm o} -\{ j=1, 2,  \ldots, i-1\}$.

 \begin{table}[htbp]
  \centering
  \caption{Payoff of player $i$.}
\begin{tabular}{|c|c|c|} \hline
\multicolumn{2}{|c|}{} &player $i'$ \\ \hline
\multicolumn{2}{|c|}{} & $i'$ \\ \hline
&$1$& $0$ \\ \cline{2-3}
&$\vdots$& $\cdots$ \\ \cline{2-3}
player $i$&$i-1$& $0$ \\ \cline{2-3}
&$i$&$u_{ii}$   \\ \cline{2-3}
&$\vdots$&$\cdots$  \\ \cline{2-3}
&$n$&$u_{in}$  \\ \hline
    \end{tabular}
  \label{Tab:PAYOFFi}%
\end{table}%
\end{proof}
\clearpage

\end{document}